%% file: neurips_2026.tex
\documentclass{article}

\usepackage[main, final]{neurips_2026}

\usepackage[utf8]{inputenc} % allow utf-8 input
\usepackage[T1]{fontenc}    % use 8-bit T1 fonts
\usepackage{hyperref}       % hyperlinks
\usepackage{url}            % simple URL typesetting
\usepackage{booktabs}       % professional-quality tables
\usepackage{amsfonts}       % blackboard math symbols
\usepackage{amsmath}
\usepackage{amssymb}
\usepackage{nicefrac}       % compact symbols for 1/2, etc.
\usepackage{microtype}      % microtypography
\usepackage{xcolor}         % colors
\usepackage{graphicx}
\usepackage{wrapfig}
\usepackage{multirow}
\usepackage[table]{xcolor}
\usepackage{colortbl}
\usepackage{subcaption}     % for side-by-side ablation+heatmap subfigures (if used)
\usepackage{tcolorbox}
\tcbuselibrary{breakable, skins}
\usepackage{hyperref}
\hypersetup{
  colorlinks=true,
  urlcolor=blue,
  linkcolor=blue,
  citecolor=blue
}

\newcommand{\xname}{ChipMATE}

\title{\xname: \underline{M}ulti-\underline{A}gent \underline{T}raining via Reinforcement Learning for \underline{E}nhanced RTL Generation}

\author{%
  Zhongkai Yu$^{*}$\\
  UCSD\\
  La Jolla, USA \\
  \texttt{zhy055@ucsd.edu} \\
  \And 
  Yichen Lin$^{*}$\\
  UCSD\\
  La Jolla, USA \\
  \texttt{yil384@ucsd.edu} \\
  \And 
  Chenyang Zhou\\
  Columbia University\\
  New York, USA \\
  \texttt{cz2791@columbia.edu} \\
   \And 
  Yuwei Zhang\\
  UCSD\\
  La Jolla, USA \\
  \texttt{yuz163@ucsd.edu} \\
   \And 
  Kun Zhou\\
  UCSD\\
  La Jolla, USA \\
  \texttt{kuzhou@ucsd.edu} \\
  \And
  Junxia Cui\\
  UCSD\\
  La Jolla, USA \\
  \texttt{jucui@ucsd.edu} \\
  \And
  Haotian Ye\\
  UCSD\\
  La Jolla, USA \\
  \texttt{h5ye@ucsd.edu} \\
  \And
  Zhengding Hu\\
  UCSD\\
  La Jolla, USA \\
  \texttt{zhh068@ucsd.edu} \\
  \And
  Zaifeng Pan\\
  UCSD\\
  La Jolla, USA \\
  \texttt{zapan@ucsd.edu} \\
  \And
  Ruiyi Wang\\
  UCSD\\
  La Jolla, USA \\
  \texttt{ruw079@ucsd.edu} \\
  \And
  Yujie Zhao\\
  UCSD\\
  La Jolla, USA \\
  \texttt{yuz285@ucsd.edu} \\
  \And
  Hejia Zhang\\
  UCSD\\
  La Jolla, USA \\
  \texttt{hez024@ucsd.edu} \\
  \And
  Jingbo Shang\\
  UCSD\\
  La Jolla, USA \\
  \texttt{jshang@ucsd.edu} \\
  \And
  Jishen Zhao\\
  UCSD\\
  La Jolla, USA \\
  \texttt{jzhao@ucsd.edu} \\
  \And
  Yufei Ding\\
  UCSD\\
  La Jolla, USA \\
  \texttt{yufeiding@ucsd.edu} \\
}

\begin{document}

\renewcommand{\thefootnote}{}
\footnotetext{$^{*}$Equal contribution.}

\makeatletter
\renewcommand{\@notice}{}
\makeatother

\maketitle

\begin{abstract}
\input{src/abstract}
\end{abstract}

% \vspace{0.5em}
% \noindent\textbf{Project page:} \url{https://github.com/zhongkaiyu/ChipMATE}

\input{src/0_intro}
\input{src/1_method}
\input{src/2_exp}
\input{src/3_related_work}
\input{src/4_conclusion}

% \newpage
\bibliography{neurips_2026}
\bibliographystyle{plain}

%%%%%%%%%%%%%%%%%%%%%%%%%%%%%%%%%%%%%%%%%%%%%%%%%%%%%%%%%%%%

\appendix

% \section{System Prompt Details}
% \label{sec:appendix-prompt}

% Full system prompts for the Verilog agent and Python reference-model agent will be provided here.
\input{src/appendix_prompts}

% \section{Additional Experimental Results}
% \label{sec:appendix-exp}

% Additional ablation studies, per-category breakdowns, and qualitative examples will be provided here.

%%%%%%%%%%%%%%%%%%%%%%%%%%%%%%%%%%%%%%%%%%%%%%%%%%%%%%%%%%%%

\newpage

\end{document}

%% file: src/abstract.tex
Existing API-based agentic systems for RTL code generation are fundamentally misaligned with industrial practice: they assume a golden testbench is available at generation time, rely on closed-source APIs incompatible with chip vendors' air-gapped security requirements, and cannot be trained on vendors' proprietary RTL codebases, leaving valuable internal data unused. Recent self-trained models address the deployment constraint but remain single-turn generators that overlook the critical role of verification in real industrial flows.

To bridge these gaps, we present ChipMATE, the first self-trained multi-agent framework for RTL generation. Inspired by industrial practice where correctness emerges from cross-comparison between independently written RTL modules and reference models, ChipMATE pairs a Verilog agent with a Python reference-model agent that mutually verify each other's outputs without any golden oracle. 
We design a backtrack-based inference workflow to prevent error propagation across turns, and a two-stage training pipeline that first trains each agent individually to saturate its code-generation capability, then trains the team jointly to collaborate effectively. To support the training, we further build a hybrid data-generation framework that produces 64.4K high-quality reference model training samples. 
ChipMATE achieves 75.0\% and 80.1\% pass@1 on VerilogEval V2 with 4B and 9B base models, outperforming all existing self-trained models and even DeepSeek V4 with 1600B parameters. 
Our code and model weights are publicly available in \url{https://github.com/zhongkaiyu/ChipMATE}.

% We present ChipMATE, the first self-trained multi-agent framework for RTL generation. Inspired by industrial practice where correctness emerges from cross-comparison between independently written RTL modules and reference models, ChipMATE pairs a Verilog agent with a Python reference-model agent that mutually verify each other's outputs without any golden oracle. We design a backtracking-based inference workflow to prevent error propagation across turns, and a two-stage training pipeline that first trains each agent individually, then trains them jointly to collaborate effectively. To support reference-model agent training, we further build a hybrid data-generation framework that produces 64.4K high-quality samples. ChipMATE achieves 75.0\% and 80.1\% pass@1 on VerilogEval V2 with 4B and 9B base models, outperforming all existing self-trained models and even DeepSeek V4 with 1600B parameters. We will open-source our model weights, inference workflow, and training dataset to facilitate future research.

%% file: src/0_intro.tex
\section{Introduction}

Large Language Models (LLMs) excel at general-purpose programming but struggle with Register-Transfer Level (RTL) code generation, a cornerstone of modern chip design, largely due to the scarcity of high-quality RTL data in public corpora.
To improve RTL generation quality, recent agentic workflows like MAGE~\citep{zhao2024mage} and VerilogCoder~\cite{ho2025verilogcoder} orchestrate LLMs through task decomposition and iterative self-correction, achieving promising results on academic benchmarks.

Despite these advances, current agentic RTL pipelines are fundamentally misaligned with industrial chip-design practice in three coupled ways, preventing their adoption by chip vendors. 
\textbf{First}, existing workflows require a golden testbench to drive self-correction, which does not hold for industrial settings.
In real production, testbenches are written by dedicated verification engineers and do not exist before the RTL is written.
Moreover, even when a testbench exists, it is often imperfect and may itself contain bugs, making it unsuitable to serve as a golden oracle.
\textbf{Second}, these workflows depend on LLM APIs, which conflict with the security posture of chip vendors. RTL code is treated as first-class intellectual property, and core development servers are routinely air-gapped to prevent any data leakage to third-party endpoints. \textbf{Third}, the API-centric design wastes chip companies' most valuable asset for LLM improvement: their internal RTL codebases. Years of production-grade RTL, far superior to any publicly available training data, remain entirely unused because API-based LLMs cannot be fine-tuned on proprietary code.

To address these constraints, recent efforts such as QiMeng-CodeV-R1~\cite{zhu2025codevr1} and RTLSeek~\citep{zhang2026rtlseek} train open-source LLMs that run on chip vendors' local servers, removing the dependency on external APIs and enabling training with in-house codebases.
Moreover, these models generate code in a single turn, eliminating the need for golden testbenches.
While this avoids the above issues, single-turn generation offers no mechanism for the model to check or correct its own output, resulting in low code accuracy.
This is unsurprising, as even a senior design engineer rarely writes correct RTL in one attempt.
Indeed, chip design industry has realized this problem and does not ensure code correctness by demanding perfection from any single engineer, but through a cross-verification workflow between design engineers and verification engineers.
Specifically, design engineers implement a module in RTL while verification engineers independently write a reference model in a high-level language (\emph{e.g.}, Python, SystemC, or C++) that predicts the RTL module's cycle-accurate behavior.
Since neither side is assumed correct, the two parties iteratively compare outputs, locate bugs, and apply fixes until the design is verified.
This industrial workflow directly inspires our work \xname{}, a multi-agent system in which one agent generates Verilog in the role of a design engineer and another generates a Python reference model in the role of a verification engineer.
The two agents work as a team, iteratively cross-verifying each other to produce high-quality RTL code.

However, building such a self-trained multi-agent workflow introduces three key challenges.
\textbf{(C1) Error propagation without a golden oracle.}
Since neither agent generates perfect code, a mismatch between the design agent and the verification agent does not reveal which side is wrong.
Naively asking one agent to ``correct'' the other can compound errors turn by turn, producing worse results than a single-model baseline.
\textbf{(C2) Weak individual capability and lack of collaboration.}
Off-the-shelf open-source models such as Qwen3.5-4B/9B achieve below 45\% accuracy on both Verilog generation and reference-model generation, even on simple benchmarks like VerilogEval V2~\cite{pinckney2024revisiting}.
Each agent's individual capability must therefore be strengthened through dedicated training before any meaningful collaboration can occur.
Beyond individual competence, the two agents must also learn to work as a team, a skill that standard single-model training does not provide.
\textbf{(C3) No training data for reference-model generation.}
Reference-model generation is a largely unexplored task with no publicly available training data.
One might hope to bootstrap a dataset by distilling from strong API-based LLMs, but even DeepSeek-R1 achieves below 20\% pass@1 when converting the Verilog-based QiMeng training set into Python reference models, making direct distillation impractical.
% The poor performance is not surprising: writing a reference model is fundamentally different from general Python programming, as it requires predicting cycle-accurate hardware behavior and following two's-complement hardware arithmetic patterns.
The poor performance stems from the unique demands of reference-model generation, which requires predicting cycle-accurate hardware behavior and adhering to two's-complement arithmetic patterns, tasks that LLMs are never exposed to during pretraining.

We present \xname{}, the first self-trained multi-agent framework for RTL generation that supports fully offline deployment. 
% \xname{} addresses challenges C1--C3 through three technical contributions, complemented by a full open-source release:

\begin{enumerate}

\item \textbf{Cross-verification multi-agent workflow (C1).}
We propose an agentic workflow in which a Verilog agent and a Python reference-model agent mutually verify each other's outputs.
To prevent error propagation across turns, we introduce a backtrack mechanism that automatically reverts to a previous turn whenever the current turn produces worse results.
To support this workflow, we build cross-language comparison tools that align Verilog and Python outputs, along with error-explanation tools that translate raw waveform mismatches into natural-language diagnostics the agents can act on.
We also design structured system prompts with code skeletons and detailed implementation guidance to deliver clear instructions and enhance model performance.
% , yielding a model-agnostic improvement of 1--5\% in first-attempt pass rate.

\item \textbf{Two-stage training pipeline (C2).} We design a two-stage training pipeline with dedicated dataset curation and reward functions. In Stage~1, we train each agent separately with SFT followed by RL to strengthen its individual capability on Verilog module and Python reference-model generation, since multi-agent collaboration requires each constituent agent to be sufficiently competent. 
In Stage~2, we introduce a multi-agent RL algorithm paired with a multi-level reward function that trains the two agents jointly, teaching them to collaborate effectively and further boosting end-to-end generation quality.

\item \textbf{Hybrid data-generation framework for reference models (C3).} To address the complete absence of reference-model training data, we develop a hybrid data-generation framework that combines agentic API-based synthesis with IR-level code conversion, transforming existing Verilog datasets into high-quality, chain-of-thought-augmented Python reference-model samples. We further apply targeted augmentation on task categories where LLM performance is weakest, ensuring balanced and comprehensive training coverage.

\item \textbf{Open-source release.} We have released all model weights for the \xname{} series (4B and 9B) and the complete multi-agent workflow implementation to facilitate future research.
\end{enumerate}

%% file: src/1_method.tex
\section{Methods}
\label{sec:method}

\begin{figure}[t]
    \centering
    \includegraphics[width=0.99\linewidth]{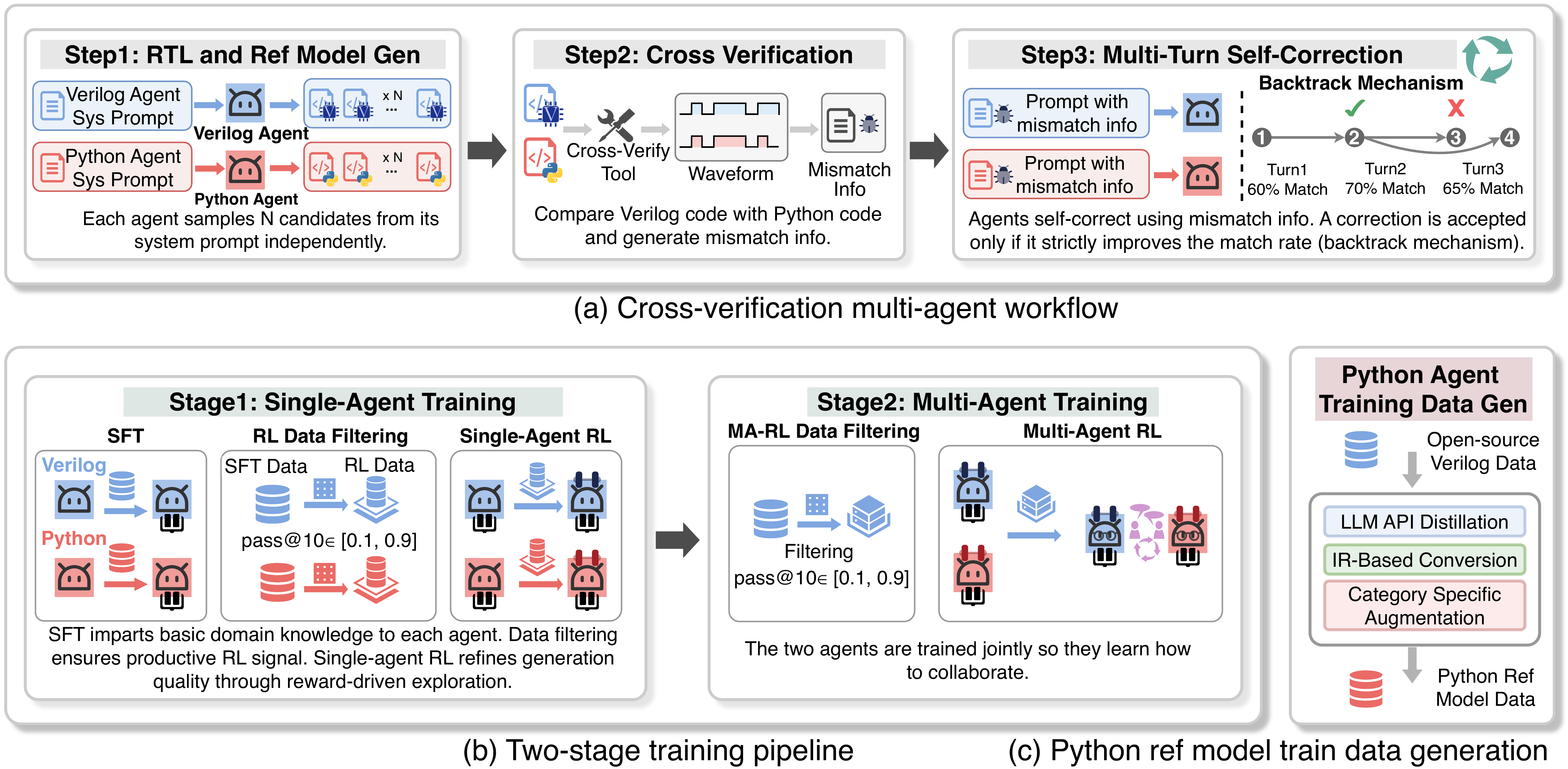}
    \caption{Overview of \xname{}. (a) \xname{} introduces multi-agent cross-verification with a backtrack mechanism.
    (b) The training pipeline of \xname{} has two stages, and we build a hybrid reference-model data generation framework.}
    \label{fig:overview}
\end{figure}

\subsection{Overview}
\label{sec:overview}
As shown in~\autoref{fig:overview}, \xname{} consists of three components: a cross-verification multi-agent workflow (\autoref{sec:workflow}) in which a Verilog agent and a Python reference-model agent iteratively refine their outputs through cross-verification, a two-stage training pipeline (\autoref{sec:training}) that first trains each agent individually and then jointly via multi-agent RL, and a hybrid data generation framework (\autoref{sec:datagen}) that creates reference-model training data from scratch.

\subsection{Cross-Verification Multi-Agent Workflow}
\label{sec:workflow}

\paragraph{Multi-turn cross-verification workflow.}
As shown in~\autoref{fig:overview}(a), our workflow proceeds in three steps. In the first step, a Verilog agent and a Python reference-model agent each sample $N$ candidate implementations from the same natural-language specification, working independently without access to each other's output. We use Python rather than SystemC as the reference-model language because open-source LLMs are substantially more proficient in it. In the second step, a cross-language comparison tool simulates both outputs on 1000 randomly generated stimuli, records waveforms, and produces structured mismatch diagnostics. In the third step, if outputs disagree, each agent receives the mismatch information and self-corrects over multiple turns. Crucially, agents never see each other's code and must independently judge whether their own implementation is at fault. To prevent error compounding across turns, we enforce a \emph{backtrack mechanism}~\citep{madaan2023selfrefine,shinn2023reflexion,chen2023selfdebug}: a correction is accepted only if it strictly improves the match rate; otherwise the agent reverts to its last accepted version. The loop terminates when outputs agree or the turn limit is reached.

\paragraph{Comparison and feedback infrastructure.}
The cross-verification tool compiles the Verilog module via Icarus Verilog and executes the Python reference model, then compares their cycle-by-cycle outputs. Since raw waveform files are difficult for LLMs to parse, we additionally build a waveform-to-natural-language converter that locates the first divergent cycle and packages the surrounding input/output context into a structured description that agents can directly understand and act upon.

\paragraph{Prompt design.}
We structure the system prompt to mirror the combinational/sequential separation found in standard RTL textbooks, since LLMs already partially internalize this decomposition from pretraining. Each prompt contains four sections: (1)~a code skeleton with the exact module name and parameterized port list, fixing the interface and preventing naming drift across rollouts. This does not constitute an unrealistic advantage, as in industrial practice the input and output signals of every module are strictly defined in specification documents and can be directly converted into a code skeleton. (2)~Combinational-logic guidelines covering latch avoidance and full case coverage; (3)~sequential-logic guidelines covering reset handling and non-blocking assignment conventions; and (4)a few-shot example demonstrating the desired reasoning-then-code format. 
This structured prompt alone improves first-attempt pass rate by approximately 1-5\% when applied to frontier LLMs via API, confirming it as a model-agnostic source of gain. Full prompt text is provided in~\autoref{app:prompts}.

\begin{figure}[t]
    \centering
    \includegraphics[width=0.99\linewidth]{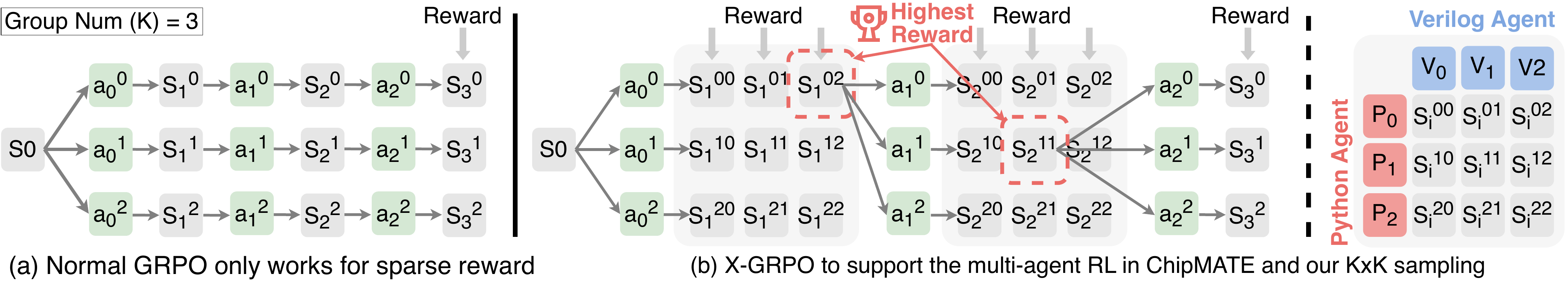}
    \caption{(a) Standard GRPO collapses to group size 1 across multi-turn rollouts when dense per-turn rewards are used. (b) X-GRPO, extending AT-GRPO~\citep{zhao2025strongertogether}, restores meaningful group variance: both agents independently sample $K$ candidates per turn, paired index-wise into $K \times K$ candidate pairs, of which the best-scoring pair is selected as the shared prefix for the next turn.}
    \label{fig:xgrpo}
\end{figure}

\subsection{Two-Stage Training Pipeline}
\label{sec:training}

As shown in~\autoref{fig:overview}, we propose a two-stage pipeline: the first stage trains each agent independently to maximize individual competence, and the second stage trains both agents jointly to learn collaborative behavior within the cross-verification workflow.

\paragraph{Stage 1: Single-agent training.}
We train the Verilog agent and the Python agent separately on their respective datasets using SFT followed by RL. SFT teaches each agent general knowledge about its target task and enables correct code generation, but with high variance, as evidenced by a large gap between pass@1 and pass@10. To close this gap, we apply GRPO~\citep{shao2024deepseekmath}, which samples a group of \(K\) candidate outputs for each query and derives group-relative advantages to guide policy optimization. We carefully curate the RL training set by retaining only problems whose post-SFT pass@10 lies in \([0.1, 0.9]\), ensuring that each group produces a productive mix of successes and failures. Problems that are always solved or never solved yield zero-variance advantages and provide no learning signal.

We adopt full-parameter fine-tuning rather than LoRA~\citep{hu2021lora} for both SFT and RL. This choice is critical because multi-agent RL in the second stage can only succeed if both participating agents are already strong. If either agent generates consistently poor code, the cross-verification loop produces noisy rewards that prevent meaningful collaboration learning.

\paragraph{Stage 2: Multi-agent RL.}
After Stage~1, both agents already excel at their respective code-generation tasks. Stage~2 teaches them to collaborate within the cross-verification workflow. Specifically, when a mismatch arises, each agent must analyze the diagnostic information, determine whether the fault lies in its own code or the other agent's, and apply self-correction only when it identifies a genuine error in its own implementation. We introduce two techniques for this stage.

\textbf{X-GRPO trajectory sampling.}
Standard GRPO is ill-suited to a multi-turn, multi-agent setting~\citep{lowe2017multiagent}.
As illustrated in~\autoref{fig:xgrpo}(a), assigning a dense reward at the end
of each turn causes the effective group size for every turn beyond the first to
collapse to one, making group-relative advantage estimation degenerate.
To address this, we draw inspiration from
Tree-of-Thought~\cite{yao2023treeofthrought} and
AT-GRPO~\citep{zhao2025strongertogether} and propose X-GRPO, a variant of GRPO that restores meaningful within-group variance across turns.
Concretely, at each turn, both agents independently generate \(K\) candidates
outputs, forming \(K \times K\) candidate pairs as the evaluation group.
After evaluating all pairs, we select the best-performing one as the shared
prefix for the next turn and repeat the process.
Because all candidates at a given turn share the same prefix, each agent's
\(K\) outputs constitute a proper GRPO group that exhibits sufficient reward
variance, keeping the advantage estimation informative throughout the entire
trajectory.
Formally, for a group \(g\) at turn \(t\), we sample \(K\) candidate actions \(\{a_t^{(c)}\}_{c=1}^K\), evaluate each with a rule-based reward \(R(\cdot)\), and compute a mean-centered, normalized advantage \(A_g(a_t^{(c)})\). Each agent has its own policy \(\theta^{(i)}\), trained on a minibatch \(\mathcal{B}_i\) that pools only that agent's groups. The clipped surrogate loss is
\begin{equation}
\label{eq:xgrpo-loss}
\mathcal{L}(\theta^{(i)}) = -\,\mathbb{E}_{g\in\mathcal{B}_i}\!\left[\frac{1}{K}\sum_{c=1}^K \min\!\Big(r_g^{(c,i)}\,A_g^{(c)},\; \mathrm{clip} \big(r_g^{(c,i)},\,1{-}\varepsilon,\,1{+}\varepsilon\big)\,A_g^{(c)}\Big)\right],
\end{equation}
where \(r_g^{(c,i)} = \pi_{\theta^{(i)}}(a_g^{(c)}\!\mid\!o_g)\,/\,\pi_{\theta_{\text{old}}^{(i)}}(a_g^{(c)}\!\mid\!o_g)\) is the importance ratio. This formulation separates the two policies' gradients while still letting them collaborate at rollout time.

\textbf{Hierarchical reward design.}
Our reward design directly serves the ultimate training goal: teaching both agents to learn from mismatch information and generate better code through multi-turn iteration. 
The per-agent reward \(R_a^{(t)}\) is composed of three components.
\textbf{(1)~Local reward} (\(R_{\text{local}}\)), which independently evaluates each agent's code correctness through turn-over-turn improvement. It follows a multi-tiered design where the score \(s \in \{0,\, 0.1,\, 0.2,\, 0.2 + 0.8c\}\) corresponds to compile failure, runtime error, I/O port mismatch, and partial pass rate \(c\), respectively. 
An agent receives the next tier of reward only after clearing all previous tiers, encouraging syntactically correct, high-pass-rate code. 
\textbf{(2)~Correct-fix bonus} (\(R_{\text{fix}}\)), a sparse binary reward for successfully resolving a previously mismatched stimulus. It is awarded only when a mismatch from the previous turn is fixed, and both agents produce the correct output at that stimulus cycle. If the two agents agree but their shared output is wrong, this reward is withheld. This teaches agents to fix mismatches correctly rather than to merely converge on an arbitrary answer.
\textbf{(3)~Team-match reward} (\(R_{\text{match}}\)), a dense reward proportional to the overall match ratio between the two agents' outputs, providing a smooth gradient toward mutual agreement.
The aggregate reward is
\begin{equation}
\label{eq:xgrpo-reward}
R_a^{(t)} = \delta_{\text{local}} \cdot R_{\text{local},a}^{(t)} + \delta_{\text{fix}} \cdot R_{\text{fix}}^{(t)} + \delta_{\text{match}} \cdot R_{\text{match}}^{(t)},
\end{equation}
where we set \(\delta_{\text{local}} = 10\), \(\delta_{\text{fix}} = 0.2\), and \(\delta_{\text{match}} = 0.5\). The large \(\delta_{\text{local}}\) makes individual code improvement the dominant training signal, the binary \(\delta_{\text{fix}}\) provides rare but high-quality debugging reward, and the moderate \(\delta_{\text{match}}\) acts as a dense regularizer that steers both agents toward agreement without dominating the optimization.

\subsection{Hybrid Reference-Model Data Generation}
\label{sec:datagen}

\begin{wrapfigure}{l}{0.6\linewidth}
    \centering
    \includegraphics[width=\linewidth]{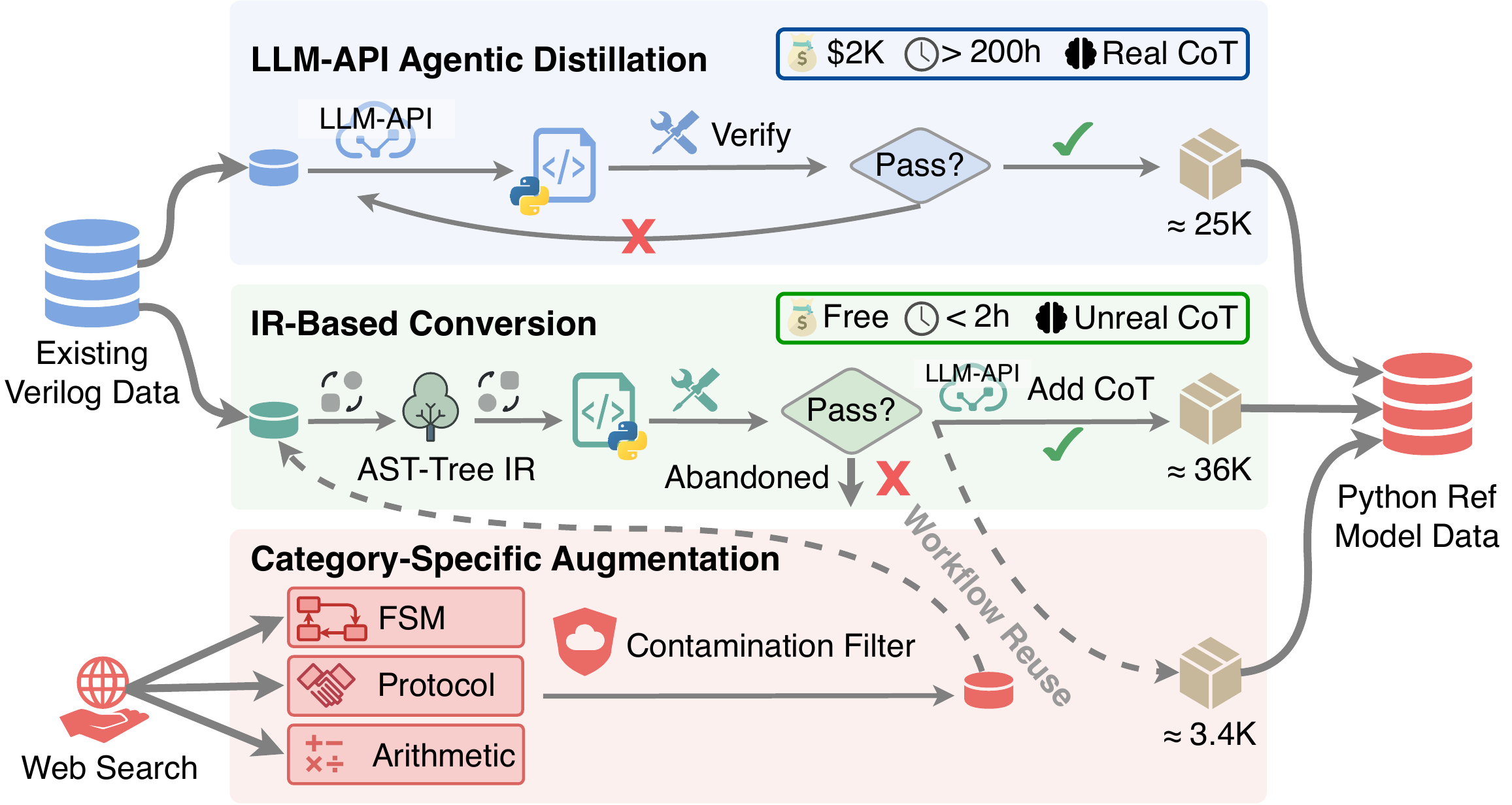}
    \caption{Hybrid data generation framework combining LLM-API distillation, IR-based conversion, and category-specific augmentation to produce the reference-model training corpus.}
    \label{fig:datagen}
\end{wrapfigure}

The two-stage training pipeline requires substantial data for both agents. Existing public datasets with chain-of-thought annotations already meet the need for Verilog agent training, but no comparable resource exists for the Python reference-model agent. To bridge this gap, we build a hybrid generation framework combining three complementary pipelines, as illustrated in~\autoref{fig:datagen}.

\paragraph{LLM-API agentic distillation.}
We call a frontier LLM to generate a Python reference model for each Verilog module, then verify it against the golden Verilog with our cross-language comparator. When mismatches are detected, diagnostic information is fed back for iterative self-correction. This pipeline produces samples with CoT reasoning, but suffers from high cost and low yield: on the 87K-sample QiMeng-CodeV~\citep{zhu2025codevr1} dataset, DeepSeek-R1~\citep{guo2025deepseekr1} achieves only 17\% single-turn and 35\% two-turn success rates. Processing the full dataset with four API keys in parallel costs over \$2{,}000 and more than 200 hours, ultimately producing only about 25K verified samples.

% \begin{figure}[t]
%     \centering
%     \includegraphics[width=0.7\linewidth]{fig/3_data_gen.pdf}
%     \caption{Hybrid data generation framework combining LLM-API distillation, IR-based conversion, and category-specific augmentation to produce the reference-model training corpus.}
%     \label{fig:datagen}
% \end{figure}

\paragraph{IR-based conversion.}
Since 25K samples alone are insufficient to train a strong model, we introduce a deterministic conversion pipeline for the remaining 62K samples.
The low LLM success rate likely arises because LLMs are never explicitly trained to replicate Verilog semantics in Python.
Our pipeline sidesteps this difficulty by operating at the abstract syntax tree (AST) level, reserving LLM calls solely for appending chain-of-thought~\citep{wei2022chainofthought} annotations to already-correct code.
The pipeline proceeds in three stages.
\textbf{(1)~Parsing.}
We parse the reference Verilog with \textsc{PyVerilog}~\citep{takamaeda2015pyverilog} and normalize its structures, including ports, signals, continuous assignments, combinational and sequential blocks into a uniform intermediate representation~(IR).
\textbf{(2)~Behavioral lowering.}
Each combinational block is converted into a pure Python function
whose output depends solely on its current inputs.
Each sequential block, triggered on \texttt{posedge clk},
becomes a per-cycle update function that applies explicit bit-width masks to preserve two's-complement semantics.
\textbf{(3)~Top-module wrapping.}
The translated functions are assembled into a \texttt{TopModule} class whose port interface mirrors the original Verilog module, enabling it to be directly tested with our cross-verification tool.
% {\color{red}Round-trip verification on the full training set achieves a 96.7\% stimulus-level pass rate; the remaining 3.3\% of samples are automatically discarded.}
This pipeline is cost-free, completes in under two hours, and yields 36K additional verified samples. 
Its principal limitation is the unreal CoT appended by LLM after the Python module is generated and verified.

\paragraph{Category-specific augmentation.}
The two pipelines above yield approximately 61K samples, sufficient in quantity but not in coverage. 
After training Qwen3.5-4B on this dataset and evaluating on VerilogEval, we find that most failures concentrate in three categories:
finite-state machines, multi-cycle protocol blocks (e.g.\ counters and FIFOs),
and bit-level arithmetic.
We collect additional Verilog examples targeting these categories via web search, convert them with the IR-based conversion pipeline and apply a contamination filter to exclude any overlap with benchmark test cases.
% that rejects any candidate matching a test problem ID or exceeding 0.3 Jaccard similarity on 5-grams of normalized port lists and module identifiers. 
This adds approximately 3.4K targeted samples, shifting the three categories' share from approximately 15\% to 28\% and strengthening the model on precisely the task types where it is weakest.

%% file: src/2_exp.tex
\section{Experiments}
\label{sec:experiments}

\subsection{Implementation Details}
\label{sec:impl_details}

% \paragraph{Training pipeline.}
We obtain \xname{} by applying our two-stage training pipeline
(\autoref{sec:training}) to the Qwen3.5~\cite{qwen35} base model (4B and 9B).
Stage~1 comprises single-agent SFT followed by single-agent RL, and Stage~2
performs multi-agent RL.
For SFT, we use LLaMA-Factory~\cite{zheng2024llamafactory} to fine-tune
Qwen3.5. The Verilog agent is trained on the 87K-sample dataset released with
QiMeng CodeV-R1~\cite{zhu2025codevr1}, and the Python agent is trained on
64.4K reference-model samples generated by our hybrid framework
(\autoref{sec:datagen}). Both agents are trained for 6~epochs with a learning
rate of $1\!\times\!10^{-5}$, a global batch size of~64, and a context length
of 16{,}384 tokens.
For single-agent RL, we use verl~\cite{sheng2024verl} with
GRPO~\cite{shao2024deepseekmath}. We set the global batch size and minibatch
size to 128, the GRPO group size to 16, and the learning rate to
$1\!\times\!10^{-6}$. Training runs for 300~steps with a rollout temperature
of 1.0, an instruction length of 2{,}048, and a response length of 16{,}384.
The reward function is the same local reward used in multi-agent RL.
For multi-agent RL, we build on a modified version of
PettingLLMs~\cite{zhao2025strongertogether} with verl as the backbone. Both
agents are updated for an additional 200~steps at the same learning rate. The
9B model is trained on two nodes of 8$\times$H200 (141\,GB each), and the 4B
model on a single 8$\times$H100 node.

% \paragraph{Evaluation protocol.}
We evaluate \xname{} on four widely used RTL generation benchmarks:
VerilogEval~v2~\cite{pinckney2024revisiting},
RTLLM~v2~\cite{liu2024openllmrtl}, ChipBench-SC (the self-contained subset of ChipBench (78 cases)~\cite{yu2026chipbench}), and CVDP~cid03~\cite{pinckney2025cvdp}.
The maximum context length is 16{,}384 tokens for all benchmarks. The generation temperature is set to 0.6 for the SFT-only checkpoint and 1.0 for
the +\,RL and agentic checkpoints. We generate 10 responses per query and
report pass@$k$~\citep{chen2021codex} for $k\!\in\!\{1,5\}$.

\input{table/table1_e2e}

%--------------------------------------------------------------------
\subsection{End-to-End Results}
\label{sec:e2e}

\autoref{tab:main} reports pass@$k$ on the four benchmarks against
large-scale foundation LLMs (GPT, Claude, and DeepSeek),
hardware-specialized baselines (CodeV-R1~\cite{zhu2025codevr1}), and
open-source base models. The top-3 results in each column are highlighted with
gold, silver, and bronze cells.

Compared with the previous state-of-the-art self-trained model, QiMeng
CodeV-R1, \xname{}-Agents-9B achieves 6.7\%--13.6\% higher pass@1. Even the
smaller \xname{}-Agents-4B improves over CodeV-R1 by 3.3\%--6.6\% despite
having fewer parameters (4B vs.\ 7B). When compared with state-of-the-art
API-based LLMs that are 20--1000$\times$ larger, \xname{}-Agents-9B outperforms all of
them. The only comparable model is Claude Opus~4.7 and GPT-5.5, which achieves better
performance on VerilogEval and CVDP but falls behind \xname{} on RTLLM v2 and
ChipBench.

Notably, the gap between pass@5 and pass@1 for \xname{} is considerably
smaller than for other models. This is because the ~\xname{} already
self-corrects errors during inference, increasing the likelihood of producing
correct code on the first attempt. This property aligns well with industrial needs, where pass@1 is the primary concern because chip design tolerates
very few bugs.

%--------------------------------------------------------------------
\subsection{Python Reference Model Generation}
\label{sec:py_ref}

\autoref{tab:python} reports the results for Python reference-model
generation. \xname{}-Python-9B and \xname{}-Python-4B both rank among the top two across all benchmarks, which we attribute to the quality of our synthetic dataset and training pipeline. 
Simulating hardware behaviour in Python is a fundamentally different task from general-purpose Python programming, and dedicated supervision turns out to be essential. 
Notably, \xname{}-Python-9B achieves 5.4--15.2\% higher pass@1 on the Python
track than \xname{}-Verilog-9B does on the Verilog track across all four
benchmarks (+7.1\% on VerilogEval~v2, +5.4\% on RTLLM~v2, +13.3\% on ChipBench-SC, and +15.2\% on CVDP). 
This suggests that LLMs can readily leverage pre-trained Python priors once they receive targeted fine-tuning on hardware-behaviour simulation, which
is precisely what our dataset and pipeline are designed to provide.

Interestingly, foundation models such as DeepSeek score lower on the Python
reference-model task than on direct Verilog generation. Given their
well-established strength in general Python coding, we believe even lightweight fine-tuning on reference-model generation could substantially improve their
hardware-design performance.

%--------------------------------------------------------------------
\subsection{Python vs.\ Verilog}
\label{sec:py_vs_v}

\begin{figure}[t]
    \centering
    \includegraphics[width=0.99\linewidth]{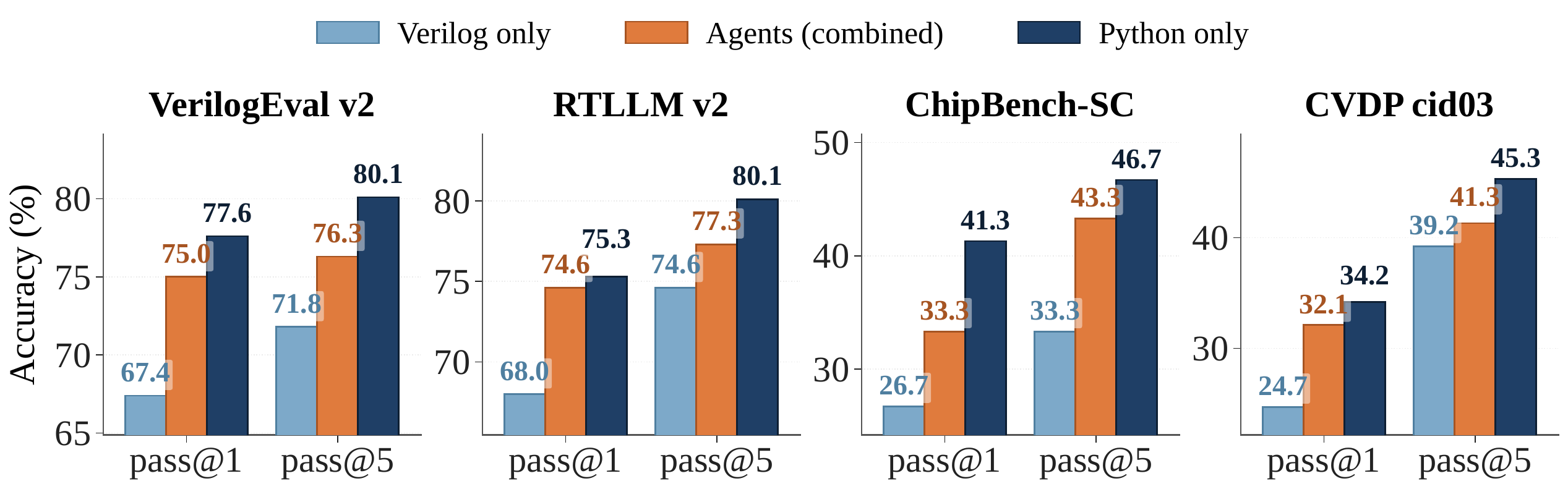}
    \caption{Pass@$k$ of different variant of \xname{}-4B. The agentic workflow consistently lands between the two single agents.}
    \label{fig:py_vs_v}
\end{figure}

To understand how our multi-agent workflow improves Verilog generation, we
compare the three \xname{}-4B variants in \autoref{fig:py_vs_v}. The Python
agent consistently achieves the highest accuracy while the Verilog agent shows
the lowest, with the multi-agent workflow landing between the two across all
benchmarks. This indicates that the workflow effectively lifts Verilog
generation quality toward the level of the stronger Python agent. It also
highlights the critical role of reference-model generation: its accuracy sets the upper bound for the entire multi-agent workflow, yet this task has been
largely overlooked by prior work.

% \input{table/tab2_python_ref_generation}
\input{table/table2_python}

\subsection{Ablation Study}
\label{sec:ablation}
\autoref{fig:ablation} presents the ablation study on VerilogEval~v2 pass@1, tracing the contribution of each technique and training stage. Among all stages, SFT yields the largest gain (+19.2\% on 4B, +22.0\% on 9B), as it equips the LLMs with foundational Verilog generation knowledge. Single-agent RL and multi-agent RL then contribute a further 1.4\%--6.5\% improvement.

Notably, simply applying a multi-agent workflow without backtracking leads to an unexpected drop of 11.6\%--14.3\%. We attribute this to the fact that neither agent produces perfect code, and naively accepting every intermediate result causes severe error propagation. Once the backtracking mechanism is introduced, however, performance rebounds by 16\%--18.6\%, surpassing the single-agent result by 4.3\%--4.4\% and demonstrating the effectiveness of multi-agent collaboration.

\subsection{Exploration of the Agentic Workflow}
\label{sec:workflow_explore}
\begin{figure}[t]
    \centering
    \begin{minipage}[t]{0.46\linewidth}
        \centering
        \includegraphics[width=\linewidth]{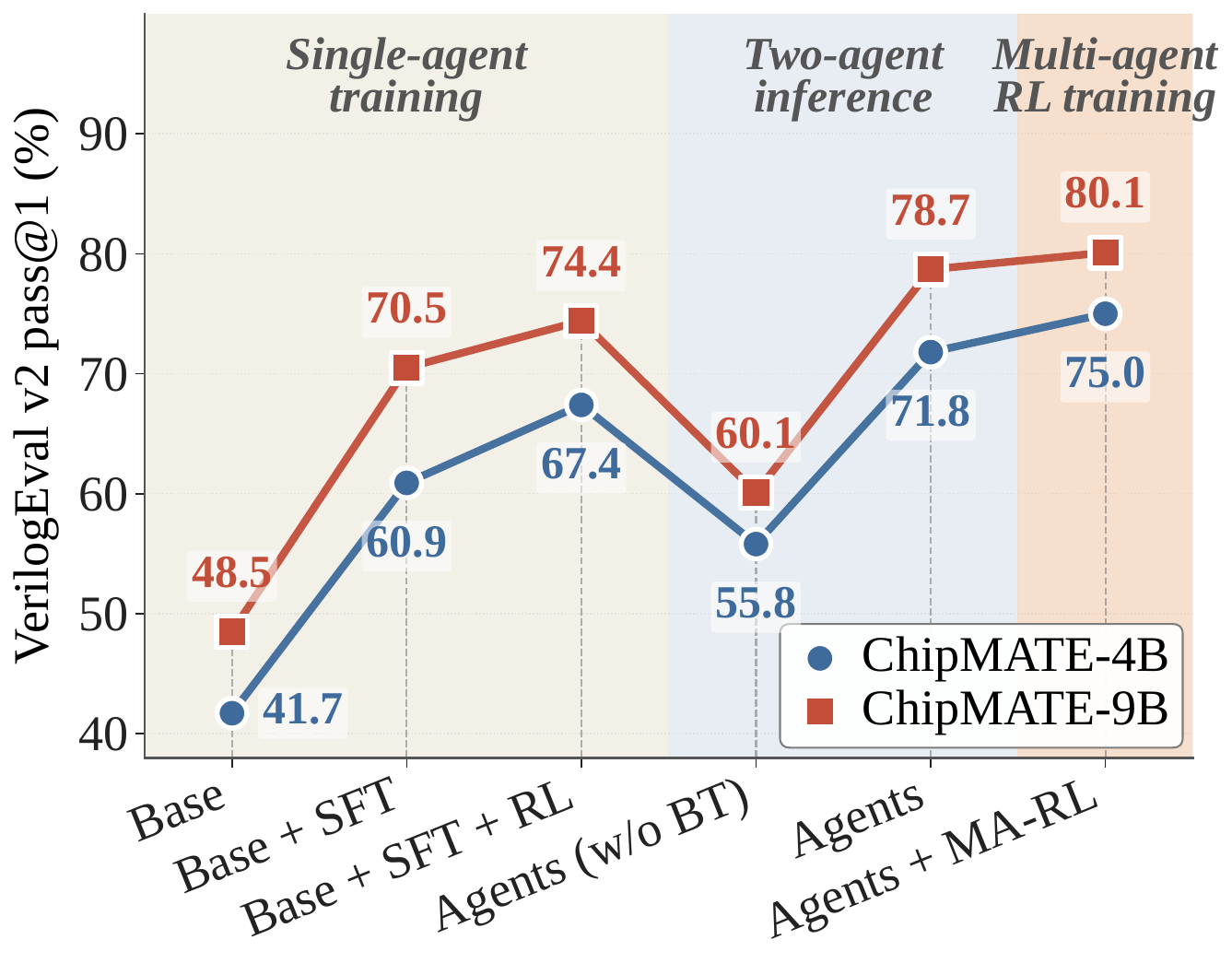}
        \caption{Ablation study on VerilogEval~v2 pass@1, showing the contribution of each technique and training step. (BT: Backtracking)}
        \label{fig:ablation}
    \end{minipage}%
    \hfill
    \begin{minipage}[t]{0.50\linewidth}
        \centering
        \includegraphics[width=\linewidth]{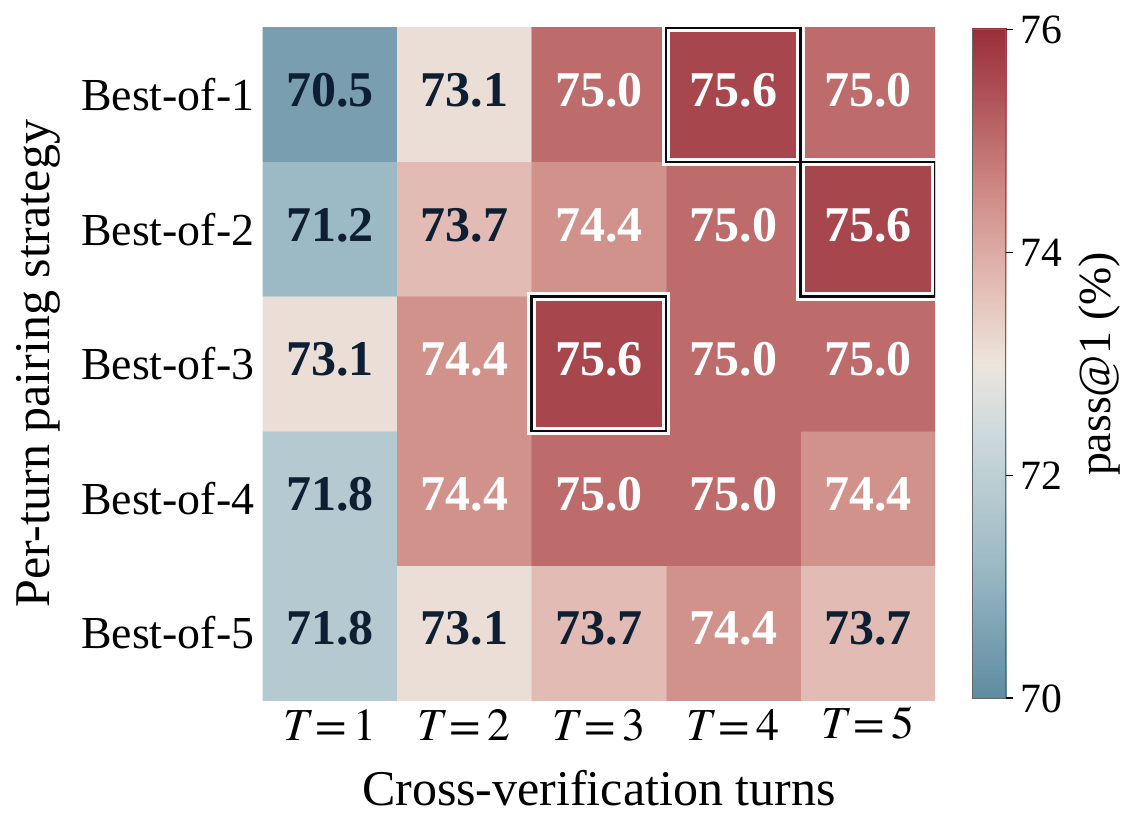}
        \caption{Exploration of multi-agent parameters using \xname{}-4B on VerilogEval pass@1. The combination of 3 turns and 3 samples is optimal.}
        \label{fig:workflow}
    \end{minipage}
\end{figure}

We finally study the design space of the agentic workflow by fixing the \xname{}-4B checkpoint and sweeping the per-turn sampling budget (Best-of-$N$, $N\!\in\!\{1,\dots,5\}$) and the maximum number of turns ($T\!\in\!\{1,\dots,5\}$). 
As shown in \autoref{fig:workflow}, accuracy does not increase monotonically with either factor, and a moderate setting of $N\!=\!3$, $T\!=\!3$ achieves the best result (75.6 pass@1). We therefore adopt \emph{Best-of-3, $T\!=\!3$} as the default configuration, as it reaches the peak accuracy at minimal inference cost. 
% We hope this finding offers useful guidance for future multi-agent workflow design.

% We use VerilogEval~v2 alone for this sweep---the trends are
% stable across benchmarks (Section~\ref{sec:e2e}), and a single
% benchmark keeps the search tractable. Two findings emerge.

% \textbf{(1)~More candidate pairs is not better.} Pass@1 peaks
% at \emph{Best-of-3} ($N\!=\!3$) and degrades at larger $N$;
% larger candidate pools enlarge the cross-verifier's selection
% task and make it easier to lock onto a plausible-but-incorrect
% pair. The Cartesian \emph{Best-of-$N\!\times\!N$} variants
% (Appendix~\ref{app:full_workflow}) are uniformly worse than
% their \emph{Best-of-$N$} counterparts at every $(N,T)$ pair.

% \textbf{(2)~Three turns are sufficient.} Across every row,
% accuracy plateaus by $T\!=\!3$. Additional turns yield
% diminishing or even negative returns (e.g., \emph{Best-of-3}
% drops 0.6 from $T\!=\!3$ to $T\!=\!4$).

%--------------------------------------------------------------------
% Combined figure for §3.5 (Ablation) and §3.6 (Workflow exploration),
% laid out side-by-side using minipage (no extra package required).
% Each minipage has its own \caption{} so each subfigure is numbered as
% an independent \autoref-able figure.

%% file: table/table1_e2e.tex
% =====================================================================
%  Main results table — NeurIPS style with rotated category column
%  Required packages (add to your preamble):
%     \usepackage{booktabs}
%     \usepackage{multirow}
%     \usepackage[table]{xcolor}
%     \usepackage{colortbl}
%     \usepackage{graphicx}   % for \resizebox & \rotatebox
% =====================================================================

% ---- Top-3 ranking colors (gold / silver / bronze) -------------------
\definecolor{rankone}{HTML}{FFD966}
\definecolor{ranktwo}{HTML}{D9D9D9}
\definecolor{rankthree}{HTML}{E8B98A}
\definecolor{ourrow}{HTML}{EEF3F8}

% ---- Ranking macros --------------------------------------------------
\providecommand{\first}[1]{\cellcolor{rankone}\textbf{#1}}
\providecommand{\second}[1]{\cellcolor{ranktwo}\underline{#1}}
\providecommand{\third}[1]{\cellcolor{rankthree}#1}

\begin{table}[t]
\centering
\caption{Pass@$k$ (\%) for Verilog generation on four benchmarks.
Within each column,
\colorbox{rankone}{\textbf{1st}}(Gold),
\colorbox{ranktwo}{\underline{2nd}}(Silver), and
\colorbox{rankthree}{3rd}(Bronze)
are highlighted.}
\label{tab:main}
\setlength{\tabcolsep}{4.5pt}
\renewcommand{\arraystretch}{1.10}
\resizebox{\textwidth}{!}{%
\begin{tabular}{@{}c l c cc cc cc cc@{}}
\toprule
\multirow{2}{*}{\textbf{Type}} & \multirow{2}{*}{\textbf{Model}} & \multirow{2}{*}{\textbf{Size}}
 & \multicolumn{2}{c}{\textbf{VerilogEval v2}}
 & \multicolumn{2}{c}{\textbf{RTLLM v2}}
 & \multicolumn{2}{c}{\textbf{ChipBench-SC}}
 & \multicolumn{2}{c}{\textbf{CVDP cid03}} \\
\cmidrule(lr){4-5}\cmidrule(lr){6-7}\cmidrule(lr){8-9}\cmidrule(lr){10-11}
 & & & \textsc{p@1} & \textsc{p@5}
   & \textsc{p@1} & \textsc{p@5}
   & \textsc{p@1} & \textsc{p@5}
   & \textsc{p@1} & \textsc{p@5} \\
\midrule
\multirow{6}{*}{\shortstack[c]{Foundation \\ Models}}
 & GPT-4o            & --   & 64.1          & 73.7          & 56.5          & 70.3          & 20.0          & 33.3          & 39.0          & 40.4          \\
 & GPT-5.5           & --   & \second{84.7} & \first{90.4}  & 63.2          & 68.0          & 30.7          & 36.7          & \first{44.0}  & \first{48.7}  \\
 & Claude Opus 4.7   & --   & \first{86.9}  & \first{90.4}  & 64.8          & 68.0          & \third{31.3}  & \first{46.7}  & \second{42.8} & \second{47.9} \\
 & DeepSeek Coder    & 236B & 68.5          & 80.8          & 57.6          & 70.0          & 16.7          & 30.0          & 22.3          & 37.2          \\
 & DeepSeek V4       & 1.6T & 67.3          & 80.1          & 58.8          & 66.0          & 18.0          & 36.7          & 21.5          & 34.6          \\
 & DeepSeek R1       & 671B & 77.5          & \second{84.7} & 64.7          & \third{75.8}  & 26.7          & \third{40.0}  & 27.7          & 42.1          \\
\midrule
\multirow{2}{*}{\shortstack[c]{Specialized \\ Models}}
 & CodeV-R1 (distill)& 7B   & 65.2          & 75.2          & 57.2          & 71.9          & 13.3          & 26.7          & 26.2          & 42.1          \\
 & CodeV-R1          & 7B   & 68.8          & 78.2          & 68.0          & \first{78.2}  & 30.0          & \third{40.0}  & 26.8          & 43.3          \\
\midrule
\multirow{2}{*}{\shortstack[c]{Base \\ Models}}
 & Qwen3.5-4B        & 4B   & 41.7          & 60.9          & 34.3          & 49.7          & 6.7           & 10.0          & 11.8          & 13.9          \\
 & Qwen3.5-9B        & 9B   & 48.5          & 66.6          & 36.1          & 57.8          & 13.3          & 20.0          & 13.3          & 21.5          \\
\midrule
\cellcolor{ourrow} & \cellcolor{ourrow}\xname{}-Verilog-4B & \cellcolor{ourrow}4B & \cellcolor{ourrow}67.4 & \cellcolor{ourrow}71.8 & \cellcolor{ourrow}68.0 & \cellcolor{ourrow}74.6 & \cellcolor{ourrow}26.7 & \cellcolor{ourrow}33.3 & \cellcolor{ourrow}24.7 & \cellcolor{ourrow}39.2 \\
\cellcolor{ourrow} & \cellcolor{ourrow}\xname{}-Agents-4B  & \cellcolor{ourrow}4B & \cellcolor{ourrow}75.0 & \cellcolor{ourrow}76.3 & \second{74.6} & \second{77.3} & \second{33.3} & \second{43.3} & \cellcolor{ourrow}32.1 & \cellcolor{ourrow}41.3 \\
\cellcolor{ourrow} & \cellcolor{ourrow}\xname{}-Verilog-9B & \cellcolor{ourrow}9B & \cellcolor{ourrow}75.3 & \cellcolor{ourrow}77.6 & \third{71.9}  & \third{75.8}  & \cellcolor{ourrow}30.0 & \cellcolor{ourrow}36.7 & \cellcolor{ourrow}28.1 & \cellcolor{ourrow}42.1 \\
\multirow{-4}{*}{\cellcolor{ourrow}\shortstack[c]{ChipMATE \\ (Ours)}}
 & \cellcolor{ourrow}\xname{}-Agents-9B  & \cellcolor{ourrow}9B & \third{80.1}  & \third{82.4}  & \first{75.8}  & \second{77.3} & \first{36.7}  & \second{43.3} & \third{40.4}  & \third{44.6}  \\
\bottomrule
\end{tabular}}
\end{table}

%% file: table/table2_python.tex
% =====================================================================
%  Python-track results table — same NeurIPS style as table1.tex
%  Required packages (add to your preamble):
%     \usepackage{booktabs}
%     \usepackage{multirow}
%     \usepackage[table]{xcolor}
%     \usepackage{colortbl}
%     \usepackage{graphicx}   % for \resizebox & \rotatebox
% =====================================================================

% ---- Top-3 ranking colors (idempotent if table1.tex already loaded) --
\definecolor{rankone}{HTML}{FFD966}
\definecolor{ranktwo}{HTML}{D9D9D9}
\definecolor{rankthree}{HTML}{E8B98A}
\definecolor{ourrow}{HTML}{EEF3F8}

% ---- Ranking macros (no-op if already defined by table1) -------------
\providecommand{\first}[1]{\cellcolor{rankone}\textbf{#1}}
\providecommand{\second}[1]{\cellcolor{ranktwo}\underline{#1}}
\providecommand{\third}[1]{\cellcolor{rankthree}#1}

\begin{table}[t]
\centering
\caption{Pass@$k$ accuracy (\%) of the \textbf{Python reference-model track}
on four hardware-code benchmarks.
Within each column,
\colorbox{rankone}{\textbf{1st}},
\colorbox{ranktwo}{\underline{2nd}}, and
\colorbox{rankthree}{3rd}
are highlighted (ties share a rank).
ChipMate-Python (ours) rows are shaded.}
\label{tab:python}
\setlength{\tabcolsep}{4.5pt}
\renewcommand{\arraystretch}{1.40}
\resizebox{\textwidth}{!}{%
\begin{tabular}{@{}c l c cc cc cc cc@{}}
\toprule
\multirow{2}{*}{\textbf{Type}} & \multirow{2}{*}{\textbf{Model}} & \multirow{2}{*}{\textbf{Size}}
 & \multicolumn{2}{c}{\textbf{VerilogEval v2}}
 & \multicolumn{2}{c}{\textbf{RTLLM v2}}
 & \multicolumn{2}{c}{\textbf{ChipBench-SC}}
 & \multicolumn{2}{c}{\textbf{CVDP cid03}} \\
\cmidrule(lr){4-5}\cmidrule(lr){6-7}\cmidrule(lr){8-9}\cmidrule(lr){10-11}
 & & & \textsc{p@1} & \textsc{p@5}
   & \textsc{p@1} & \textsc{p@5}
   & \textsc{p@1} & \textsc{p@5}
   & \textsc{p@1} & \textsc{p@5} \\
\midrule
\multirow{4}{*}{\shortstack[c]{Foundation \\ Models}}
 & GPT-5.5              & --   & \third{61.5}  & \third{75.5}  & 48.5          & 55.4          & \third{33.0}  & \third{42.1}  & \second{41.6} & \first{47.4}  \\
 & DeepSeek Coder       & 236B & 60.1          & 73.1          & 42.4          & 50.0          & 28.7          & 40.0          & 21.5          & 35.1          \\
 & DeepSeek V4          & 1.6T & 59.7          & 71.8          & 44.4          & 54.3          & 30.7          & 40.0          & 24.7          & 36.5          \\
 & DeepSeek R1          & 671B & 57.1          & 70.7          & \third{49.6}  & \third{57.8}  & 28.7          & 36.7          & 26.2          & 37.2          \\
\midrule
\multirow{2}{*}{\shortstack[c]{Specialized \\ Models}}
 & CodeV-R1 (distill)   & 7B   & 40.4          & 48.5          & 36.0          & 46.2          & 23.3          & 30.0          & 24.7          & 37.2          \\
 & CodeV-R1             & 7B   & 45.3          & 56.4          & 42.7          & 51.8          & 28.7          & 33.3          & 22.3          & 34.6          \\
\midrule
\multirow{2}{*}{\shortstack[c]{Base \\ Models}}
 & Qwen3.5-4B           & 4B   & 46.8          & 63.7          & 38.2          & 46.1          & 16.7          & 23.3          & 10.2          & 12.7          \\
 & Qwen3.5-9B           & 9B   & 48.0          & 64.7          & 40.1          & 51.8          & 20.0          & 26.7          & 11.4          & 19.9          \\
\midrule
\cellcolor{ourrow} & \cellcolor{ourrow}ChipMate-Python-4B & \cellcolor{ourrow}4B & \second{77.6} & \second{80.1} & \second{75.3} & \second{80.1} & \second{41.3} & \second{46.7} & \third{34.2}  & \third{45.3}  \\
\multirow{-2}{*}{\cellcolor{ourrow}\shortstack[c]{ChipMate-P \\ (Ours)}}
 & \cellcolor{ourrow}ChipMate-Python-9B & \cellcolor{ourrow}9B & \first{82.4}  & \first{84.7}  & \first{77.3}  & \first{81.0}  & \first{43.3}  & \first{50.0}  & \first{43.3}  & \second{46.7} \\
\bottomrule
\end{tabular}}
\end{table}

%% file: src/3_related_work.tex
\section{Related Work}
\label{sec:related}

\paragraph{API-based agentic Verilog generation.}
A large body of work improves RTL generation by wrapping frontier closed-source models in
agentic workflows~\citep{thakur2023autochip,ho2025verilogcoder,islam2024aivril,mi2024coopetitivev,hong2023metagpt,qian2023chatdev,yang2024sweagent}.
MAGE~\citep{zhao2024mage} decomposes the pipeline into specialized agents for the testbench
generation, candidate sampling, and state-checkpoint debugging.
VerilogCoder~\citep{ho2025verilogcoder} further adds graph-based task planning and AST-driven
waveform analysis.
All of these systems require continuous access to proprietary API endpoints, conflicting with
the security posture of chip vendors that air-gap core design servers, and offer no path to
training on a company's internal RTL codebase.

\paragraph{Self-trained models for RTL generation.}
A parallel line trains locally deployable models via supervised fine-tuning
(RTLCoder~\citep{liu2025rtlcoder}, AutoVCoder~\citep{gao2024autovcoder},
BetterV~\citep{pei2024betterv}, OriGen~\citep{he2024origen}, CraftRTL~\citep{liu2024craftrtl})
or reinforcement learning with verifiable rewards
(VeriReason~\citep{wang2025verireason}, RTLSeek~\citep{zhang2026rtlseek}, ChipSeek-R1~\citep{chipseekr1}).
QiMeng-CodeV-R1~\citep{zhu2025codevr1} combines rule-based testbench generation,
round-trip data synthesis, and adaptive DAPO to train a strong local Verilog model.
RTLSeek~\citep{zhang2026rtlseek} optimizes design correctness and diversity through multi-stage
diversity-oriented RL.
These works address the deployment limitations of API-based systems but all rely on a golden
testbench or reference implementation as an oracle, and model only the design engineer's role.

Concurrent to our work, SiliconMind-V1~\citep{chen2026siliconmind} locally deploys a
self-trained multi-agent workflow for Verilog generation.
\xname{} differs in that it trains two independent models (Verilog and Python reference-model
agents) and develops dedicated training data for reference-model generation, a task where even
671B frontier models achieve below 20\% pass@1, enabling code-blind cross-verification that
mirrors industrial design-verification practice.

%% file: src/4_conclusion.tex
\section{Conclusion}
\label{sec:conclusion}

We presented \xname{}, a self-trained multi-agent framework for RTL code generation that supports fully offline deployment. 
\xname{} pairs a Verilog agent with a Python reference-model agent that cross-verify outputs without any golden testbench or cloud-LLM API. 
Three contributions underpin the design: a backtracking-based workflow that enforces monotonic quality improvement across turns, a two-stage training pipeline tailored for multi-agent collaboration, and a hybrid data-generation framework that produces high-quality reference model training data. Through this work, we aim to demonstrate the potential of self-trained multi-agent workflows for Verilog generation, offering a paradigm better aligned with industrial practice where data privacy is a top priority.

% emphasis the important of reference model generation, and we hope 

% Experiments on standard benchmarks confirm state-of-the-art RTL accuracy among locally deployable
% systems. We hope \xname{} opens a path toward chip design automation that is secure, customizable,
% and continuously improvable, with verification treated as a trained first-class participant rather
% than an external oracle.

%% file: src/appendix_prompts.tex
% =============================================================================
%  Appendix: Agent Prompts
%  All prompts taken verbatim from the V31/V32 SFT data
%  (data/codev_r1_sft_v32_combined.jsonl,
%   data/codev_r1_TESTSET_FOR_NEXT_PHASE_python_v2.jsonl).
%  Required preamble (place once in main file):
%      \usepackage{tcolorbox}
%      \tcbuselibrary{breakable, skins}
% =============================================================================
\section{Agent Prompt Templates}
\label{app:prompts}

We list the system and user-message templates used by the Verilog agent
and the Python reference-model agent during single-turn SFT and multi-agent
rollouts, together with the retry-and-fix prompt fragments that X-GRPO
injects on turns $t\!>\!0$ whenever a V/P mismatch is detected. Both agents
share a unified
\texttt{\textless think\textgreater...\textless /think\textgreater\textless answer\textgreater...\textless /answer\textgreater}
reasoning format.

\subsection{Verilog Agent System Prompt}
\label{app:prompt-v}

\begin{tcolorbox}[breakable,colback=gray!5,colframe=gray!50,
                  boxrule=0.4pt,arc=2pt,
                  fonttitle=\bfseries,title=Verilog agent system prompt]
\small\ttfamily\raggedright
You are a helpful assistant. The assistant first thinks about the
reasoning process in the mind and then provides the user with the answer.
The reasoning process and answer are enclosed within \textless think\textgreater\
\textless /think\textgreater\ and \textless answer\textgreater\
\textless /answer\textgreater\ tags, respectively, i.e.,
\textless think\textgreater\ reasoning process here
\textless /think\textgreater\textless answer\textgreater\ answer here
\textless /answer\textgreater. Now the user asks you to write verilog code.
After thinking, when you finally reach a conclusion, enclose the final
verilog code in \texttt{```verilog ```} within \textless answer\textgreater\
\textless /answer\textgreater\ tags. i.e., \textless answer\textgreater\
\texttt{```verilog\textbackslash n module top\_module(in, out, ...) ... ```}
\textless /answer\textgreater.
\end{tcolorbox}

\paragraph{Example user message.}
The user message gives the natural-language behaviour and a structured
port list (no Verilog code skeleton---only the interface):
\begin{tcolorbox}[breakable,colback=gray!5,colframe=gray!50,
                  boxrule=0.4pt,arc=2pt,fonttitle=\bfseries,
                  title={Example user message (Verilog agent, VerilogEval-style)}]
\small\ttfamily\raggedright
I would like you to implement a module named TopModule with the following
interface. All input and output ports are one bit unless otherwise
specified.\\[0.4em]
\hspace*{1em}- input\hspace*{1em}in\hspace*{1em}(3 bits)\\
\hspace*{1em}- output out (2 bits)\\[0.4em]
The module should implement a "population count" circuit that counts the
number of '1's in the input vector.
\end{tcolorbox}

\subsection{Python Reference-Model Agent System Prompt}
\label{app:prompt-p}

The Python agent operates in a \emph{skeleton-completion} mode: the user
message contains a natural-language spec \emph{plus} a Python class
skeleton that pins the class name, method signature, and exact signal
names / bit-widths. The agent must complete the bodies of
\texttt{\_\_init\_\_} and \texttt{eval} without renaming any identifier so
that the cross-language verifier can drive the resulting class identically
to the Verilog DUT.

\begin{tcolorbox}[breakable,colback=gray!5,colframe=gray!50,
                  boxrule=0.4pt,arc=2pt,
                  fonttitle=\bfseries,title=Python reference-model agent system prompt]
\small\ttfamily\raggedright
You are an expert hardware modeling assistant. Your task is to write
Python code that serves as a cycle-accurate reference model for a Verilog
hardware module. Your Python code's outputs must match the expected
outputs of the Verilog circuit exactly.\\[0.4em]
The user will provide a natural language description of the module's
behavior and a Python code skeleton. You must complete the provided
skeleton. Do not rename the class, method, or any signals. Use the exact
signal names and bit-widths from the skeleton.\\[0.4em]
GENERAL RULES:\\
\hspace*{1em}- Use bitwise operators: \&\ \textbar\ \textasciicircum\ \textasciitilde\
\textgreater\textgreater\ \textless\textless. Each \texttt{eval()} call =
one rising clock edge. Never read clk.\\
\hspace*{1em}- Mask all outputs and state variables to their correct
bit-width, e.g.\ \texttt{val \& ((1 \textless\textless\ N) - 1)}.\\[0.4em]
FOR COMBINATIONAL CIRCUITS (no reset, no state):\\
\hspace*{1em}- Compute outputs directly from inputs in \texttt{eval()}. No
state variables needed.\\[0.4em]
FOR SEQUENTIAL CIRCUITS (has reset and/or state):\\
\hspace*{1em}- Declare all state variables in \texttt{\_\_init\_\_()},
initialized to 0.\\
\hspace*{1em}- In \texttt{eval()}: if reset is asserted, set all state and
outputs to 0 and return.\\
\hspace*{1em}- Otherwise: compute next-state from current state
(\texttt{self.*}) and inputs, compute outputs, then update
\texttt{self.*} at the end.\\[0.4em]
First, think step-by-step inside \textless think\textgreater\
\textless /think\textgreater\ tags. Then, provide your final Python code
inside \textless answer\textgreater\ \textless /answer\textgreater\ tags
with \texttt{```python ```} markers.
\end{tcolorbox}

\paragraph{Example user message (with skeleton).}
The user prompt injects the spec, a signal table, and a pre-populated
\texttt{class TopModule} skeleton in which the state variables and the
input bit-masks are already laid out; the agent only fills in the
sequential / combinational logic and the output expression.

\begin{tcolorbox}[breakable,colback=gray!5,colframe=gray!50,
                  boxrule=0.4pt,arc=2pt,fonttitle=\bfseries,
                  title={Example user message (Python agent, 3-stage pipelined ALU)}]
\small\ttfamily\raggedright
Design a pipelined ALU that performs a sequence of arithmetic operations
in three stages. The ALU should take four 10-bit inputs (\texttt{a},
\texttt{b}, \texttt{c}, \texttt{d}) and two clock signals
(\texttt{clk1}, \texttt{clk2}) and produce a 10-bit output (\texttt{F}).
The operations are as follows:\\
\hspace*{1em}1. \textbf{Stage 1}: Compute \texttt{a + b} and \texttt{c - d},
and store the value of \texttt{d}.\\
\hspace*{1em}2. \textbf{Stage 2}: Compute the sum of the results from
Stage 1.\\
\hspace*{1em}3. \textbf{Stage 3}: Compute the product of the result from
Stage 2 and the stored value of \texttt{d}.\\[0.4em]
The ALU should use pipelining to ensure that each stage operates
independently, and the final result should be available at the output
\texttt{F}. This Python class, named \texttt{alu\_op}, has the interface
designed in a port table (signal name, direction, width, description) for
each of \texttt{a}, \texttt{b}, \texttt{c}, \texttt{d}, \texttt{clk1},
\texttt{clk2}, and \texttt{F}.\\[0.4em]
Complete this skeleton:\\
\texttt{```python}\\
\texttt{class TopModule:}\\
\hspace*{1em}\texttt{def \_\_init\_\_(self):}\\
\hspace*{2em}\texttt{self.stage1\_d\hspace*{6pt}= 0}\\
\hspace*{2em}\texttt{self.stage1\_diff\hspace*{2pt}= 0}\\
\hspace*{2em}\texttt{self.stage1\_sum\hspace*{4pt}= 0}\\
\hspace*{2em}\texttt{self.stage2\_sum\hspace*{4pt}= 0}\\
\hspace*{2em}\texttt{self.stage3\_product = 0}\\[0.2em]
\hspace*{1em}\texttt{def eval(self, inputs: dict) -\textgreater\ dict:}\\
\hspace*{2em}\texttt{a = inputs.get("a", 0) \& 0x3FF}\\
\hspace*{2em}\texttt{b = inputs.get("b", 0) \& 0x3FF}\\
\hspace*{2em}\texttt{c = inputs.get("c", 0) \& 0x3FF}\\
\hspace*{2em}\texttt{d = inputs.get("d", 0) \& 0x3FF}\\
\hspace*{2em}\texttt{\# TODO: implement sequential logic}\\
\hspace*{2em}\texttt{return \{"F": ...\}}\\
\texttt{```}\\[0.4em]
Answer:
\end{tcolorbox}

\subsection{Retry-and-Fix Prompt Fragments (X-GRPO turns $t\!>\!0$)}
\label{app:prompt-retry}

When the previous turn's V/P pair produced any mismatch on the
random-stimuli cross-verification, X-GRPO appends three fragments to the
\emph{user message} of the next turn for the chosen best pair: previous
code attempts (\texttt{\{previous\_code\}}), the diff log
(\texttt{\{previous\_error\_log\}}), and a one-sentence refinement
instruction (\texttt{\{refine\_instr\}}). On $t\!=\!0$ all three are empty
strings, recovering the single-turn behaviour shown above. We display the
Verilog-side fragments; the Python-side is structurally identical
(substitute \texttt{verilog}$\to$\texttt{python},
\texttt{module code}$\to$\texttt{class TopModule},
\texttt{//}$\to$\texttt{\#}).

\paragraph{(a) Previous-code fragment \texttt{\{previous\_code\}}.}
\label{app:prompt-prevcode}
The two most recent valid attempts (truncated to 1500 chars each) are
re-injected so the agent can see its own trajectory:
\begin{tcolorbox}[breakable,colback=blue!3,colframe=blue!30,
                  boxrule=0.4pt,arc=2pt,fonttitle=\bfseries,
                  title={Previous-code fragment (Verilog side)}]
\small\ttfamily\raggedright
Your previous Verilog attempts:\\
Attempt $k\!-\!1$:\\
\texttt{```verilog}\\
\textit{\textless truncated previous code, $\le$ 1500 chars\textgreater}\\
\texttt{// ...(code truncated)...}\\
\texttt{```}\\[0.3em]
Attempt $k$:\\
\texttt{```verilog}\\
\textit{\textless truncated previous code, $\le$ 1500 chars\textgreater}\\
\texttt{// ...(code truncated)...}\\
\texttt{```}
\end{tcolorbox}

\paragraph{(b) Verification-error fragment \texttt{\{previous\_error\_log\}}.}
\label{app:prompt-errlog}
A structured diff of the cross-verifier's per-stimulus mismatches
(truncated to 2000 chars). Crucially, the message frames the disagreement
as \emph{mutual}: the agent is told its peer might be the one at fault,
preventing the V agent from over-correcting when the Python reference is
actually wrong.
\begin{tcolorbox}[breakable,colback=blue!3,colframe=blue!30,
                  boxrule=0.4pt,arc=2pt,fonttitle=\bfseries,
                  title={Verification-error fragment (Verilog side)}]
\small\ttfamily\raggedright
Previous verification error:\\
Verilog vs Python: $m$/$N$ mismatches across $T$ test vectors.\\
First mismatches (got = your Verilog, exp = peer Python):\\
\hspace*{1em}Test 0, signal `out': got=42, exp=37\\
\hspace*{2em}(inputs: a=5, b=3, clk=1)\\
\hspace*{1em}Test 4, signal `out': got=12, exp=8\\
\hspace*{2em}(inputs: a=2, b=6, clk=1)\\
\hspace*{1em}\textit{...(up to 5 mismatches shown)...}\\
Check your logic carefully. Either you or the Python agent is wrong ---\\
only change your code if you think your previous code is wrong.
\end{tcolorbox}

\paragraph{(c) Refinement instruction \texttt{\{refine\_instr\}}.}
\label{app:prompt-refine}
A short, fixed instruction emitted only on $t\!>\!0$ turns:
\begin{tcolorbox}[breakable,colback=blue!3,colframe=blue!30,
                  boxrule=0.4pt,arc=2pt,fonttitle=\bfseries,
                  title={Refinement instruction}]
\small\ttfamily\raggedright
\textbf{(Verilog side)}\\
Please refine your Verilog code to improve correctness and quality. You
MUST output the complete module code in \texttt{```verilog```} blocks.\\[0.4em]
\textbf{(Python side)}\\
Please refine your Python code based on the mismatch feedback above.
Output complete \texttt{class TopModule} in \texttt{```python```} blocks.
\end{tcolorbox}